\documentclass[prl,twocolumn,amsmath,amssymb,superscriptaddress,preprints]{revtex4}
\usepackage{graphicx}
\def\be{\begin{equation}}       \def\ee{\end{equation}}
\def\bea{\begin{eqnarray}}      \def\eea{\end{eqnarray}}

\begin{document}

\title{Magnetic Properties of the Superconducting State of Iron-Based Superconductors}

\author{Kangjun Seo}
%\email{ seo@physics.purdue.edu }
\affiliation{Department of Physics, Purdue University, West Lafayette, Indiana 47907, USA}

\author{Chen Fang}
%\email{ cfang@purdue.edu }
\affiliation{Department of Physics, Purdue University, West Lafayette, Indiana 47907, USA}
\author{B. Andrei Bernevig}
\affiliation{Princeton Center for Theoretical Science, Princeton
University, NJ 08544}
\author{Jiangping Hu}
%\email{ hu4@physics.purdue.edu}
\affiliation{Department of Physics, Purdue University, West Lafayette, Indiana 47907, USA}

\date{\today}

\begin{abstract}
We show that features of the dynamical spin susceptibility can
unambiguously  distinguish between different pairing symmetries of
the superconducting state in iron-based superconductors. A magnetic
resonance is a unique feature for the extend $s_{x^2y^2}$-wave $\cos
k_x\cos k_y$
 pairing symmetry. It is present in the pure superconducting (SC) state, but weakens in the
 mixed SC and magnetically ordered state. We calculate the the RPA correction to the NMR spin relaxation rate $1/T_1$ and the Knight
shift in the above states and show a good agreement with experimental
results.
Moreover, we argue that the energy dispersion of the magnetic
resonance along c axis observed in neutron scattering experiments is
also an indirect evidence supporting the $s_{x^2y^2}$ pairing
symmetry.
\end{abstract}

\maketitle

%%%%%%%%%%%%%%%%%%%%%%%%%%%%%%%%%%%%%%%%%%%%%%%%%%%%%%%%%%%%%%%%%%

%%%%%%%%%%%%%%%%%%%%%%%%%%%%%%%%%%%%%%%%%%%%%%%%%%%%%%%%%%%%%%%%%%

\paragraph*{Introduction --}
The recently discovered iron-based  high-temperature
superconductors\cite{kamihara2008,takahashi2008,
GFchen2008,Chenxh2008,wen2008} have very intriguing magnetic
properties\cite{mook} which may hold the key to understanding their
superconducting pairing mechanism. Theoretically, many possible gap
pairing symmetries have been proposed. Due to the proximity of the
superconducting state to a collinear antiferromagnetic state, as
well as due to relatively weak phonon interactions, a
magnetism-based superconducting mechanism is favored.  Falsifying or
verifying the prediction of pairing symmetry will be a significant
step in understanding the nature of the  magnetically driven SC
pairing mechanism.

Both weak and strong coupling approaches
suggest that an extended $s$-wave pairing symmetry is
 favored \cite{seo2008,Mazin2008a,Wang2009}. Based on a weak-coupling
approach, an $s$-wave, (so called $s^{\pm}$) state\cite{Mazin2008a}
in which the sign of order parameters changes between
hole and electron pockets, is argued to be favored for repulsive inter-band interactions. However, the
weak-coupling approach does not specify the exact form of order
parameters, and is dependent on the degree of nesting between the electron and hole pockets. In the strong coupling approach, in a recent paper
\cite{seo2008}, we showed that the pairing symmetry is determined
mainly by the \emph{next} nearest neighbor antiferromagnetic
exchange coupling $J_2$\cite{Fang2008,Yildirim2008,si} and has an
explicit form in momentum space, $\cos(k_x)\cdot \cos(k_y)$. This
result is model-independent, \emph{as long as}
the dominating interaction is next-nearest neighbor $J_2$
\emph{and} the Fermi surfaces are located close to the $\Gamma$
and $M$ points in the Brillouin zone. The $\cos(k_x) \cdot
\cos(k_y)$ changes sign between the electron and hole pockets in
the Brillouin zone. In this sense, it resembles the order
parameter, $s^{\pm}$, proposed through weak-coupling general
arguments \cite{Mazin2008a}.

The magnitudes of superconducting gaps measured by ARPES at
different Fermi surfaces are in a good agreement with the simple
$|cos(k_x)cos(k_y)|$ form \cite{Ding2008a,Nakayama2008a,Hasan2008},
but directly probing the sign change from electron to hole pockets
still remains a big experimental challenge, despite several
theoretical proposals \cite{Tsai2008,Ghaemi2008,Parker2008a,Wu2009}.
Fortunately, the magnetic properties in the superconducting state
may provide us with indirect evidence of the pairing symmetry. Maier
and Scalapino\cite{Maier2008d} first pointed out that, in the
extended $s$-wave SC states, there is a strong coherent peak in the
dynamic spin susceptibility $\chi(Q,\omega)$ at an energy $\hbar
\omega$ below the two-gap value and at a wavevector $Q=(0,\pi)$
identical to the magnetic ordering wavevector in parent compounds.
Recent neutron scattering
experiments\cite{Christianson2008,Lumsden2008,Chi2008} have observed
a magnetic resonance in the superconducting states similar to the
magnetic resonance observed in cuprates. This result provides an
indirect evidence of the extended $s$-wave pairing if the coherent
peak is a unique feature of this pairing symmetry. Neutron
scattering experiments\cite{Chi2008} have also observed a
significant dispersion of resonance peak along c axis.

Complementing neutron scattering, NMR is an important probe of magnetic
properties. The experimental results of the NMR spin relaxation rate
$1/T_1$ in these superconductors have been a challenge for theories
predicting the extended $s$-wave pairing symmetry due to the absence
of coherent peak and the near cubic power law
dependence on temperature \cite{Matano2008,Terasaki2009,Fukazawa2009,Mukuka2009}.
Several works\cite{Parish2008,Parker2008b,Laad2009} have addressed
this problem and pointed out that for the $\cos(k_x)\cdot
\cos(k_y)$ order parameter, the inter-band contribution to the NMR
spin relaxation rate does \emph{not} exhibit a coherence peak. As such, the experimental result is not inconsistent with the pairing
symmetry if the inter-band contribution is much larger than the
intra-band contribution or if the samples are strongly disordered.
However, in a simple mean-field state, the intra-band contribution
is always larger or comparable to the inter-band contribution and the spin
relaxation rate still exhibits an enhancement right below $T_c$
owing to its fully gapped $s$-wave nature\cite{Parish2008}.

In this paper, we perform a detailed calculation of the magnetic
response in the superconducting state of iron-based superconductors.
Within a two orbital model and within the RPA approximation we
address several experimentally related issues: (1) We demonstrate
that the coherent resonance peak is an unique feature of the extend
$s$-wave pairing as suggested in ref.\cite{Daghofer2008,Maier2008d}.
We show that the different pairing symmetries in iron-based
superconductors are distinguishable by the distinct features of
their corresponding dynamical spin
 susceptibilities; (2) We investigate the magnetic susceptibility in the mixed SC and SDW
  state and find that the coherent peak is strongly weakened in the mixed
state.  Many experiments suggest a possible coexistence of SC and
magnetic ordering in underdoped materials
\cite{Khasanov2008,Drew2008,Xiao2009},  and the calculated doping
dependence of the resonance peak could be checked in future
experiments; (3) We show that by considering the RPA correction to
the mean-field BCS state, both the $1/T_1$ and the Knight shift show
a good agreement with experimental results even when the intra-band
contributions are included. The RPA correction of the mean-field BCS
state can dramatically enhance the inter-band contribution to the
NMR spin relaxation time $1/T_1$; (4) We show that the dispersion of
the magnetic resonance along the c axis observed in neutron
scattering experiments\cite{Chi2008} can also be interpreted as
indirect evidence of the $s_{x^2y^2}$ pairing symmetry. The magnetic
exchange coupling along c-axis has two effects on the resonance: (1)
modifying the SC gap along c-axis and (2) causing new RPA
corrections. Although the second effect produces the right trend of
the dispersion, it is too small to explain the dispersion observed
in experiments. Therefore,  it is the first effect that causes the
dispersion. Moreover, in order to explain the experimentally
observed dispersion of the resonance peak towards lower energy from
$Q_z=0$ to $Q_z=\pi$, the superconducting order parameters at
electron and hole pockets must have opposite values.

%%%%%%%%%%%%%%%%%%%%%%%%%%%%%%%%%%%%%%%%%%%%%%%%%%%%%%%%%%%%%%%%%%%
%\section{Model}
%%%%%%%%%%%%%%%%%%%%%%%%%%%%%%%%%%%%%%%%%%%%%%%%%%%%%%%%%%%%%%%%%%%
\paragraph*{Model--} The Fe ions form a square lattice in the FeAs layer of LaFeAsO system with
As ions sitting in the center of each square plaquette of the Fe
lattice and displaced above and below the Fe plane. The
crystallographic unit cell includes two Fe and two As ions. We adopt
the two orbitals per site model proposed in Ref.\cite{raghu2008},
capturing the degeneracy of the $d_{xz}$ and $d_{yz}$ orbitals on
the Fe atoms and the general location of the Fermi surfaces. The
resulting Fermi surface consists of two hole pockets, centered
around the $\Gamma=(0,0)$ and $(\pm \pi, \pm \pi)$, and two electron
pockets, centered around $(\pm \pi, 0), (0,\pm\pi)$ in the unfolded
Brillouin Zone (the hole pocket at $(\pm \pi, \pm \pi)$ is an
artifact of the two-band model,  all the LDA calculations as well as
more complicated $5$-band models\cite{Kuroki2008} have this pocket
at the $\Gamma$ point). The single electron Hamiltonian is written
as
\begin{eqnarray}
\label{ham0}
 H_0 = \sum_{\mathbf{k}\sigma}\psi^\dagger_{\mathbf{k}\sigma}
    \left( \begin{array}{cc}
            \epsilon_x (\mathbf{k})-\mu & \epsilon_{xy}(\mathbf{k}) \\
            \epsilon_{xy}(\mathbf{k}) & \epsilon_y (\mathbf{k})-\mu
           \end{array}
    \right) \psi_{\mathbf{k}\sigma},
\end{eqnarray}
where $\psi^\dagger_{\mathbf{k}\sigma} = ( c^\dagger_{1,\mathbf{k},\sigma}, c^\dagger_{2,\mathbf{k},\sigma}) $ is the creation operator with spin-$\sigma$ in the orbitals $(1,2)=(d_{xz},d_{yz})$, $\mu$ is the chemical potential. The electronic dispersions are given by
\begin{equation}
\begin{array}{l}
\epsilon_{x}(\mathbf{k}) = -2 t_1 \cos k_x-2 t_2 \cos k_y -4t_3 \cos k_x\cos k_y , \\
\epsilon_{y}(\mathbf{k}) = -2 t_1 \cos k_y-2 t_2 \cos k_x -4t_3 \cos k_x\cos k_y, \\
\epsilon_{xy}(\mathbf{k}) =  -4t_4 \sin k_x\sin k_y.
\end{array}
\end{equation}
Choosing $t_1 = -1$, $t_2=1.3$, and $t_3=t_4=-0.85$ in electron volts, the eigenvalues of Eq.~(\ref{ham0}),
\begin{equation}
E_{\pm}(\mathbf{k}) = \frac{\epsilon_{x}+\epsilon_y}{2} -\mu \pm \sqrt{\left(\frac{\epsilon_{x}-\epsilon_y}{2}\right)^2 + \epsilon^2_{xy}},
\end{equation}
yield the two Fermi pockets aforementioned. For the half-filed system, corresponding to two electrons per site, $\mu=1.54$.

In the following we shall consider the singlet pairing between
electrons within each orbital based on $t-J_1-J_2$ model proposed
in ref.\cite{seo2008}. The symmetry of the superconducting order
parameter $\Delta(\mathbf{k})$ has two possible $d$-wave types,
$d_{x^2-y^2} =\Delta_0 (\cos k_x - \cos k_y)/2$ and
\textbf{$d_{xy}= \Delta_0 \sin{k_x}\sin{k_y}$, }and two $s$-wave
types, $s_{x^2+y^2} =\Delta_0 ( \cos k_x + \cos k_y)/2 $ and
$s_{x^2y^2}=\Delta_0 \cos k_x \cos k_y$. We neglect $d_{xy}$ and
the inter-orbital pairings:  ref.\cite{seo2008} showed
they are negligible for the case of $t-J_1-J_2$ model. The
effective mean-field Hamiltonian is then given by
\begin{equation}
\label{h_mf}
H_{\text{MF}} = \sum_{\mathbf{k}}
\Psi_\mathbf{k}^\dagger h(\mathbf{k}) \Psi_\mathbf{k},
\end{equation}
\begin{equation}
\label{h_mf_mat}
h(\mathbf{k}) = \left( \begin{array}{cccc}
\xi_x(\mathbf{k})& \Delta_1(\mathbf{k}) &\epsilon_{xy}(\mathbf{k})& 0 \\
\Delta_1^\ast(\mathbf{k})& -\xi_x(\mathbf{k}) &0 &-\epsilon_{xy}(\mathbf{k}) \\
\epsilon_{xy}(\mathbf{k}) &0& \xi_y(\mathbf{k})& \Delta_2(\mathbf{k})  \\
0& -\epsilon_{xy}(\mathbf{k})& \Delta_2^\ast(\mathbf{k}) &-\xi_x(\mathbf{k})
\end{array}\right),
\end{equation}
where $\xi_x = \epsilon_x -\mu$ and $\xi_y = \epsilon_y-\mu$, and the four-component spinor $\Psi_\mathbf{k}^\dagger = (c_{1,\mathbf{k},\uparrow}^\dagger, c_{1,\mathbf{-k},\downarrow},c_{2,\mathbf{k},\uparrow}^\dagger, c_{2,\mathbf{-k},\downarrow})$. Due to the $C_4$ symmetry of the underlying lattice, $\Delta_1(\mathbf{k}) = \Delta_2(\mathbf{k})$, except for the $d_{x^2-y^2}$ case, where $\Delta_1(\mathbf{k}) = -\Delta_2(\mathbf{k})$. The $d_{x^2-y^2}$ and $s_{x^2+y^2}$  pairing symmetries are nodal and $s_{x^2y^2}$ is nodeless for any small doping parameter. Note that $s_{x^2y^2}$ exhibits a sign change between the hole and the electron pockets, while $s_{x^2+y^2}$ does not.

%%%%%%%%%%%%%%%%%%%%%%%%%%%%%%%%%%%%%%%%%%%%%%%%%%%%%%%%%%%%%%%%%%%
%\subsection{the spin susceptibility in the mixed states}
%%%%%%%%%%%%%%%%%%%%%%%%%%%%%%%%%%%%%%%%%%%%%%%%%%%%%%%%%%%%%%%%%%%
First we consider the one-loop contribution to the spin susceptibility that include the intra-band and inter-band contributions as
\be
\chi_{ij}^{\alpha\beta}(q,i\omega) = \int_0^\beta d\tau \langle T S_i^\alpha(q,\tau) S_j^\beta(-q,0)\rangle e^{i\omega \tau}
\ee
where $S_i^\alpha (q,\tau) = \sum_k c^\dagger_{\alpha,\mathbf{k+q},\mu}(\tau)\sigma^i_{\mu\nu}c_{\alpha,\mathbf{k},\nu}(\tau)$, and $\alpha,\beta$ refer to different orbital indices. The total susceptibility is given by $\chi_0^{ij}(\mathbf{q},i\omega) = \sum_{\alpha\beta} \chi_{ij}^{\alpha\beta}(\mathbf{q},i\omega)$. Since $\chi^{+-} = (\chi^{xx}+\chi^{yy})/2 = \chi^{zz}$ for the singlet pairing, we calculate the susceptibility,
\begin{equation}
\label{chi}
\chi^{zz}_0(\mathbf{q},i\omega) = -\frac{T}{2N}\sum_{\mathbf{k},n} \text{Tr} \left[ G(\mathbf{k+q},i\omega_n+i\omega) G(\mathbf{k},i\omega_n) \right],
\end{equation}
where $G(\mathbf{k},i\omega) = (i \omega 1 - H_{\text{MF}})^{-1}$.
Introducing the band operators, $\Phi_\mathbf{k}$, such that
$\Psi_\mathbf{k} = U(\mathbf{k})\Phi_\mathbf{k}$,
the Hamiltonian, $H_{\text{MF}}$, is diagonalized with the eigenvalues, $\mathcal{E}_i (\mathbf{k})$, and the eigenvectors, $M_{\alpha \beta }^{ij}(\mathbf{k},\mathbf{q})$, and the spin susceptibility becomes
\bea\nonumber
\chi_0^{zz}(\mathbf{q},i\omega) =&& -\frac{1}{2} \sum_{\mathbf{k},\alpha,\beta} M_{\alpha  \beta }^{ij}(\mathbf{k}, \mathbf{q}) \times \\
&&\left[ \frac{f(\mathcal{E}_{j}(\mathbf{k})) - f(\mathcal{E}_{i}(\mathbf{k+q}))}{i\omega - \mathcal{E}_j(\mathbf{k})+\mathcal{E}_i(\mathbf{k+q})}
\right],
\eea
where $f(\mathcal{E}_i(\mathbf{k})) = 1/(1+\exp [\mathcal{E}_i(\mathbf{k})/T])$.
For $s$-wave symmetries, $\Delta_1 = \Delta_2$, after the analytical continuation we obtain:
\begin{widetext}
\begin{equation}
\label{sus}
\chi_0^{zz}(\mathbf{q},\omega) = -\frac{1}{2} \sum_{\mathbf{k},\mu\nu=\pm}  \left[ 1 - \frac{\Delta (\mathbf{k})\Delta(\mathbf{k+q}) +{E}_\mu(\mathbf{k}){E}_\nu(\mathbf{k+q})}{\mathcal{E}_\mu(\mathbf{k}) \mathcal{E}_\nu(\mathbf{k+q})} \right]
\frac{f(\mathcal{E}_{\mu}(\mathbf{k})) - f(-\mathcal{E}_{\nu}(\mathbf{k+q}))}{\omega + i\eta - \mathcal{E}_\mu(\mathbf{k})-\mathcal{E}_\nu(\mathbf{k+q})},
\end{equation}
\end{widetext}
where $\mathcal{E}_\pm(\mathbf{k}) = \sqrt{
E_\pm^2(\mathbf{k}) + \Delta^2(\mathbf{k}) }$.

Within RPA, the spin susceptibility
$\chi_{\text{RPA}}(\mathbf{q},\omega)$ is given in matrix form by:
\begin{equation}
\chi_{\text{RPA}}(\mathbf{q},i\omega) = [\mathbf{I} - \Gamma\chi_0(\mathbf{q},i\omega)]^{-1} \chi_0(\mathbf{q},i\omega),
\end{equation}
where $\mathbf{I}$ is a unit matrix and the vertex $\Gamma =
U\mathbf{I}$ with the value of $U$ chosen in the paramagnetic
phase. $\chi_0$ above is written in $2\times 2$ matrix form whose
entries contain the orbital contributions of Eq.~(\ref{chi}). Note
that the RPA enhancement of the spin fluctuations is determined by
the $\text{det} |\mathbf{I}-\Gamma \chi_0(\mathbf{q},i\omega) |$.
In the superconducting state the quasiparticles at the hole and
electron Fermi surfaces are connected by the
$\mathbf{Q}=(\pi,0),(0,\pi)$ wave vector. In the unfolded BZ, the
$s_{x^2y^2}$ order parameter satisfies the condition
$\Delta(\mathbf{k}) = -\Delta(\mathbf{k+Q})$. The imaginary part
of the inter-band bare susceptibility is zero for small
frequencies due to the opening of the gap and experiences a
discontinuous jump at $\omega_c=\text{min}(|\Delta(\mathbf{k})|+|\Delta(\mathbf{k+Q})|,k)$.
%This satisfies the resonance condition for the inter-band susceptibility.

%
%%%%%%%%%%%%%%%%%%%%%%%%%%%%%%%%%%%%%%%%%%%%%%%%%%%%%%%%%%%%%%%%%%%
%\section{Results}
%%%%%%%%%%%%%%%%%%%%%%%%%%%%%%%%%%%%%%%%%%%%%%%%%%%%%%%%%%%%%%%%%%%

% superconducting state
 %%%%%%%%%%%%%%%%%%%%%%%%%%%%%%%%%%%%%%%%%%%%%%%%%%%%%%%%%%%%%
\begin{figure}
\includegraphics[width=8cm]{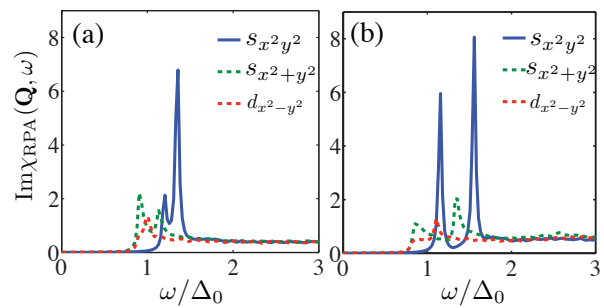}
\caption{ \label{fig01} The imaginary part of the RPA spin
susceptibility $\chi_{\text{RPA}}$ at $\mathbf{Q}=(\pi,0)$  for
various superconducting pairing symmetries. (a) hole-doped side
($\mu=1.4$) and (b) electron-doped side ($\mu=1.6$). The blue line
represents the result for $s_{x^2y^2}\sim \cos k_x \cos k_y$, the
green dashed line represents the $s_{x^2+y^2}\sim \cos k_x + \cos
k_y$, and the red dashed line represents the $d_{x^2-y^2}$ symmetry.
We used $\Delta_0 = 0.3$ and $U=0.6t_1$.  }
\end{figure}
%%%%%%%%%%%%%%%%%%%%%%%%%%%%%%%%%%%%%%%%%%%%%%%%%%%%%%%%%%%%%

\begin{figure}
\includegraphics[width=8cm]{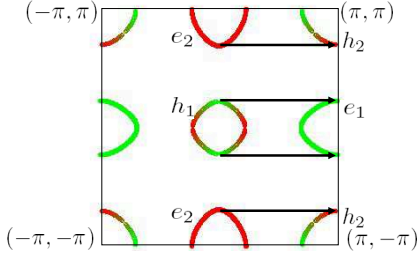}
\caption{ \label{FS} The schematic of the FSs not far away from
half-filling. The hole and electron pockets are connected by the
vector $\mathbf{Q}$.    }
\end{figure}

\paragraph*{Comparison of spin susceptibility in different pairing symmetry states:}
 Fig.~\ref{fig01} presents the imaginary part of the RPA spin
susceptibility $\text{Im}\chi_{\text{RPA}}(\mathbf{Q},\omega)$ for
various superconducting order parameters on (a) the electron-doped
side and (b) the hole-doped side. We observe
that  only the $s_{x^2y^2}$ pairing symmetry gives peaks unsuppressed by
RPA. At both electron and hole dopings, there are
two pronounced peaks for this pairing symmetry. While the number
of peaks is model-dependent and in this case characteristic to the
two-band model used, the high intensity of the peaks is a
generic feature stemming  from the coherence
factor
$(1-\Delta(\textbf{k})\Delta(\textbf{k+Q})/\mathcal{E}_+(\textbf{k})\mathcal{E}_-(\textbf{k+Q}))/2$,
where $\textbf{Q}=(\pi,0)$. This coherence factor clearly becomes
large if $\Delta(\textbf{k})=-\Delta(\textbf{k+Q})$, a condition
only met by the $s_{x^2y^2}$ pairing. The
coherence factor is weighted by the spectra weight
$\delta(\omega-\mathcal{E}_+(\textbf{k})-\mathcal{E}_-(\textbf{k+Q}))$
and integrated over $\textbf{k}$ to obtain the final $Im(\chi)$.
For the $s_{x^2y^2}$ symmetry pairing, the energy profile (after
the SC gap opening) is almost isotropic and flat around the $\Gamma$
and $X$ points. This means that at $\omega\sim\omega_{c}$,
almost all the k's around the electron and hole FSs can contribute
to the coherent peak, making its intensity significantly higher
than the other pairing symmetries. For $d_{x^2-y^2}$
($s_{x^2+y^2}$) pairing symmetries, there are four nodes on around
$\Gamma$ ($X$) point, making the energy profile around the
$\Gamma$ ($X$) point considerably angle-dependent.  Therefore it is rather difficult to find a constant $\omega_c$ such that most k's on the FSs can
contribute to the sum. Apart from the number of peaks, the previous discussion is generic
and does not depend on the particular model used. Within the two-band model, the number of peaks for the
$s_{x^2y^2}$ pairing symmetry is two. This is due to the FS
topology given by the two band model. As shown in FIG.\ref{FS} the two resonance peaks corresponds to the
nesting of two pairs of electron-hole pockets: the
nesting between $\Gamma$ hole pocket and $X$ electron pocket and
the nesting between $M$ hole pocket and $Y$ electron pocket. When
doping is small, we can roughly express the two resonance
frequencies as\begin{equation}
\begin{array}{l}
\omega_{\text{res},1}\sim|\Delta(\mathbf{k}^{h}_{1})|+|\Delta(\mathbf{k}^{e}_{1})|,\\
\omega_{\text{res},2}\sim|\Delta(\mathbf{k}^{h}_{2})|+|\Delta(\mathbf{k}^{e}_{2})|.
\end{array}
\end{equation} According to the pairing gap formula
$\cos(k_x)\cos(k_y)$, the larger FS gives a smaller gap magnitude.
Therefore we have $\omega_{\text{res},1}>\omega_{\text{res},2}$,
because the FS around $\Gamma$ is larger than that around $M$. For
the $d_{x^2-y^2}$ pairing, the two hole pockets have nodes and the
minimum gaps $\Delta(\mathbf{k}^{h}_{i})$'s on the hole surface
are hence much smaller than the electron surface gaps
$\Delta(\mathbf{k}^{e}_{i})$'s; for the $s_{x^2+y^2}$ pairing,
the two electron pockets have nodes and the gap values on the
electron pockets are much smaller. Hence, the peaks for these two
pairing symmetries shift to the lower energy than to those of the
$s_{x^2y^2}$ symmetry.

The band structure of the current two band model as the function of
doping will not exactly represent the real materials. However,   the
differences of the resonance peaks  between  electron-doped and
hole-doped  strongly suggests that the resonance peak is tied to the
combination of the opposite signs of SC order parameters and the
close matching of Fermi surfaces between electron and hole pockets.
The resonance is stronger for better nested surfaces connected by Q
vectors. This provides a direct explanation of  the orbital
selective electron-mode coupling observed in Angle Resolved
Photoemission Experiments\cite{Richard2009}.

% mixed state
%%%%%%%%%%%%%%%%%%%%%%%%%%%%%%%%%%%%%%%%%%%%%%%%%%%%%%%%%%%%%
\begin{figure}[t]
\includegraphics[width=8cm]{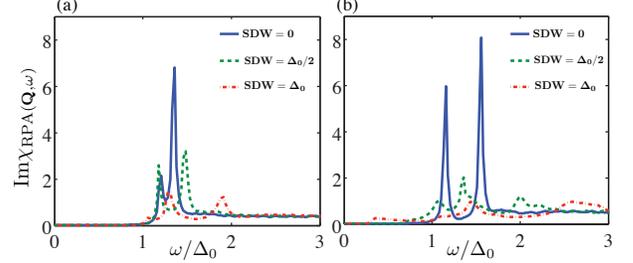}
\caption{
\label{fig02}
The imaginary part of the RPA susceptibility in the (a)hole-doped and (b)electron-doped side. We used the same $\Delta_0$ and the chemical potentials with Fig.~\ref{fig01}.
}
\end{figure}
%%%%%%%%%%%%%%%%%%%%%%%%%%%%%%%%%%%%%%%%%%%%%%%%%%%%%%%%%%%%%

\paragraph*{Magnetic resonance in the mixed SC and SDW states with $s_{x^2y^2}$
pairing symmetry} Recently, several experiments indicate the
coexistence of SC and SDW ordering in underdoped
materials\cite{Khasanov2008,Drew2008,Xiao2009}. Therefore, it is of
great interest to investigate whether the magnetic resonance
survives in the mixed state. Treating the SDW order in mean-field,
we now have off-diagonal terms between $\mathbf{k}$ and
$\mathbf{k}+\mathbf{Q_{SDW}}$, where $\mathbf{Q_{SDW}}=(\pi,0)$. In
the mean-field approximation, the spin exchange interaction,
%\begin{widetext}
\be
H_{\text{ex}}=\sum_{\mathbf{q}=(q_x,q_y)}
J(\mathbf{q})
\mathbf{S}(\mathbf{q})\cdot\mathbf{S}(-\mathbf{q}),
\ee
can be decoupled as
\be
H^{\text{MF}}_{\text{ex}}=\sum_{\mathbf{q}}J(\mathbf{q})
\langle\mathbf{S}(\mathbf{q})\rangle\cdot\mathbf{S}(-\mathbf{q})
+\text{h.c.},
\ee
%\end{widetext}
where
$J(\mathbf{q})=J_1(\cos(q_x)+\cos(q_y))+2J_2\cos(q_x)\cos(q_y)$.
Substituting the realistic magnetic structure,
$\langle\mathbf{S}(\mathbf{q})\rangle=(0,0,M)\delta_{\mathbf{q},\mathbf{Q}_{SDW}}$,
the mean-field interaction reduces to
\bea
H^{\text{MF}}_{\text{ex}}=W\sum_{i\neq j}
c^\dag_{i,\mathbf{k},\sigma}\tau^{\sigma\sigma'}_z
c_{j,\mathbf{k+Q}_{SDW},\sigma'}+\text{h.c.},
\eea
where $W=J(\mathbf{Q}_{SDW})M$ is a control parameter chosen
between $0\sim\Delta_0$ in our calculation. Therefore we can
write the full mean-field Hamiltonian as
\bea
H^{\text{MF}}=\sum_{\mathbf{k}}\mathbf{\Psi_k}^\dag\tilde{h}(\mathbf{k})\mathbf{\Psi_k},
\eea
where
\begin{widetext}
\bea
\tilde{h}(\mathbf{k})=h(\mathbf{k})\otimes(I+\tau_z)/2+h(\mathbf{k}+\mathbf{Q}_{SDW})\otimes(I-\tau_z)/2+WI\otimes{I}\otimes\tau_x,\eea
\bea
\mathbf{\Psi_{k}^\dag}=\{c^\dag_{1,\mathbf{k},\uparrow},c^\dag_{1,\mathbf{k+Q_{SDW}},\uparrow},c_{1,\mathbf{-k},\downarrow},c_{1,\mathbf{-k+Q_{SDW}},\downarrow},c^\dag_{2,\mathbf{k},\uparrow},c^\dag_{2,\mathbf{k+Q_{SDW}},\uparrow},c_{2,\mathbf{-k},\downarrow},c_{2,\mathbf{-k+Q_{SDW}},\downarrow}\},
\eea
\end{widetext}
with $I$ a 2$\times$2 identity matrix and $h(\mathbf{k})$ a 4$\times$4 matrix given in Eq.(\ref{h_mf_mat}).

Fig.~\ref{fig02} presents the RPA total spin susceptibility in the
mixed state of SDW coexisting with $s_{x^2y^2}$ order in the (a)
hole-doped  and (b) electron-doped side.   The pronounced resonance
peaks observed in the pure  $s_{x^2y^2}$ superconducting state are
reduced by the presence of the SDW. The SDW order doubles the unit
cell and reduces the original Brillouin zone which causes the weight
of the spin susceptibility at $(0,\pi)$ to spread out.

% absolute superconducting orders
% %%%%%%%%%%%%%%%%%%%%%%%%%%%%%%%%%%%%%%%%%%%%%%%%%%%%%%%%%%%%%
%\begin{figure}
%\includegraphics[width=5cm]{fig_sc_mu155s}
%\includegraphics[width=5cm]{fig_scsdw3_mu155s}
%\caption{
%\label{fig03}
%The spin susceptibility in the superconducting state with $\Delta_{sc}= 0.3 t_1$, and $\mu=1.45 t_1, 1.5 t_1$, and $1.55 t_1$, respectively.
%RED : $|\cos k_x \times\cos k_y|$, GREEN : $|\cos k_x + \cos k_y|$, and BLUE : $|\cos k_x - \cos k_y|$.}
%\end{figure}
%%%%%%%%%%%%%%%%%%%%%%%%%%%%%%%%%%%%%%%%%%%%%%%%%%%%%%%%%%%%%%

%In Fig.~\ref{fig03} the various superconducting order parameters are taken by absolute values to see the effects of the sign changes in the Brillouin zone. $S_2$ symmetric superconducting gap at the $\Gamma$ point and $(\pi,\pi)$ point have different sign. In the $S_2$-wave SC state, there is no peak observed as shown in the Fig.~\ref{fig03}a. For the mixed state we have chosen $\Delta_0 = 0.3 t_1$ and $W_{SDW}=0.3 t_1$. For absolute $S_2$-wave SC order, there shows no two peak feature observed in the $S_2$-wave SC order. And the position of the relatively broad resonance peak is located at $2W_{SDW}$ (Fig.~\ref{fig03}b).

% spin relaxation time
\paragraph*{NMR spin-relaxation rate and Knight shift}
As discussed earlier in the manuscript,  the experimental absence of
a coherent peak and the presence of near cubic power law dependence
on temperature of the NMR spin relaxation rate $1/T_1$ in these
superconductors
\cite{Matano2008,Terasaki2009,Fukazawa2009,Mukuka2009} have been a
challenge for theories predicting the extended $s$-wave pairing
symmetry. In a previous paper\cite{Parish2008}, two of us also
calculated the NMR spin relaxation rate $1/T_1$ of the bare
superconductor and found that it factorizes into inter- and
intra-band contributions. While, for the $\cos(k_x)\cdot \cos(k_y)$
order parameter, the inter-band contribution to the NMR spin
relaxation rate does \emph{not} exhibit a coherence peak, the
intra-band contribution is larger than the inter-band contribution
and still exhibits an enhancement right below $T_c$ owing to its
fully gapped $s$-wave nature. Adding the two contributions we find
that, although the coherence peak for $\cos(k_x)\cdot \cos(k_y)$ is
smaller than that for a sign-preserving gap (such as, for example,
$|\cos(k_x) \cdot \cos(k_y)|$), it is still present due to the
intra-band contribution. The coherence peak can be strongly reduced
if the intra-band scattering is stronger than inter-band scattering.
Here we also show that the NMR experimental results are consistent
with the extended $s$-wave pairing symmetry if the simple RPA
correction over the mean-field bare SC state is considered as argued
by\cite{Parker2008b}.

The spin-relaxation rate at a temperature $T$  is defined as
\begin{equation}
R = \frac{1}{T_1 T} = -\frac{1}{2\pi} \lim_{\omega_0\to 0}\frac{\text{Im}\sum_\mathbf{q} A(\mathbf{q})\chi^{+-}(\mathbf{q},\omega_0)}{\omega_0},
\end{equation}
where $A(\mathbf{q})$ is a structure factor: For F it is roughly
isotropic,  while for As, $A(\mathbf{q})=\cos q_x/2 \cos q_y/2$
(neglecting the fact that the position of As is out of plane of Fe's).
In the following we take the structure factor uniform, i.e.,
$A(\mathbf{q})=1$.
 %%%%%%%%%%%%%%%%%%%%%%%%%%%%%%%%%%%%%%%%%%%%%%%%%%%%%%%%%%%%%
\begin{figure}[t]
\includegraphics[width=8cm]{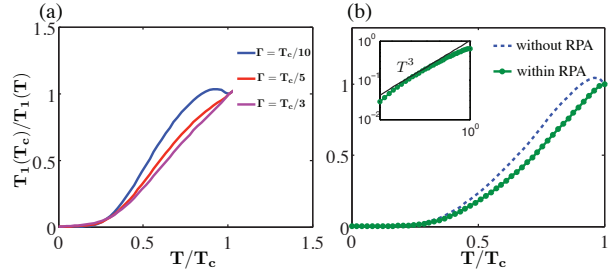}
\caption{ \label{fig03} (a)The NMR spin-relaxation time $1/T_1$ as a
function of temperature $T$  in unit of $T_c$ for various values of
$\Gamma$, which is considered as the phenomenological disorder
parameter. We used $\Delta_0/2 =T_c$ with $\Delta_0 = 0.3$. (b)The
effect of the spin fluctuation within RPA susceptibility. We choose
the on-site Coulomb repulsion $U=0.5 t_1$ such that the system
remains in the
paramagnetic state. Inset: the log-log plot of the RPA
susceptibility with line of $T^3$ showing the temperature dependence
of $\chi_{\text{RPA}}$ just below $T_c$. }
\end{figure}
%%%%%%%%%%%%%%%%%%%%%%%%%%%%%%%%%%%%%%%%%%%%%%%%%%%%%%%%%%%%%

In Fig.~\ref{fig03}, we present the NMR spin-relaxation time  as a
function of the temperature(in unit of $T_c$).   The coherence peak
is present in the bare $s_{x^2y^2}$ pairing SC state if the impurity
effect is weak, as found in \cite{Parish2008}.    The coherence peak
is reduced by the effects of disorder and the  RPA correction. The
disorder is taken in account by broadening the imaginary part of
frequency, $\Gamma$. Fig.~\ref{fig03}(a) shows that the coherence
peak at $T_c$ is reduced as disorder increases, $\Gamma =T_c/10 \sim
T_c /3$. Fig.~\ref{fig03}(b) presents the effect of the  RPA
correction. A small interaction $U\sim 0.5 t_1$  is enough to
suppress the coherence peak.

%%%%%%%%%%%%%%%%%%%%%%%%%%%%%%%%%%%%%%%%%%%%%%%%%%%%%%%%%%%%%
\begin{figure}[t]
\includegraphics[width=8cm]{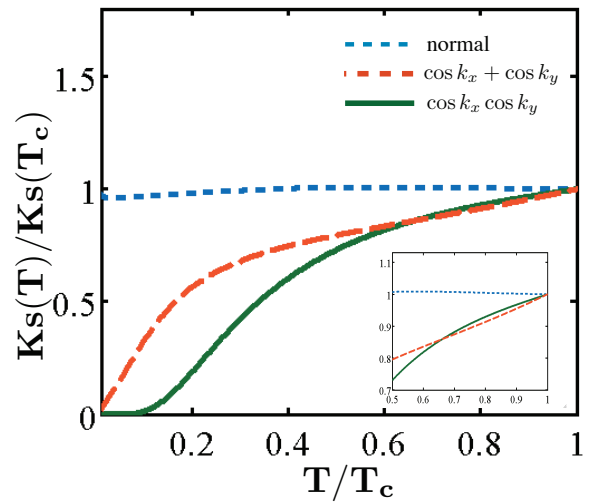}
\caption{ \label{knight_shift} Temperature dependence of the Knight
shift. The solid green line is a plot  with a sign-reversed extended
$s$-wave, $\cos k_x \cos k_y$, and the dashed red line is one with
sign-preserved extended $s$-wave, $\cos k_x + \cos k_y$. Inset shows
the difference between two order parameters just below $T_c$. }
\end{figure}
%%%%%%%%%%%%%%%%%%%%%%%%%%%%%%%%%%%%%%%%%%%%%%%%%%%%%%%%%%%%%
We also calculate the Knight shift, $K_s=\chi(\mathbf{q}\to
0,\omega=0)$.   Fig.~\ref{knight_shift} plots the Knight shift value
as a function of temperature in the different $s$-wave pairing
symmetry  states: a sign-reversed extended $s$-wave
($s_{x^2y^2}\sim\cos k_x \cos k_y$) and an extended $s$-wave
($s_{x^2+y^2}\sim\cos k_x + \cos k_y$). The main difference between
these two ordering symmetry states is that the former one is fully
gapped while the second one has nodes on the electron pockets.
At low temperatures, $K_s(T)$ has a power-law
temperature dependence for the second case. The fact that $K_s(T)$
is a concave function around $T=0.4T_C$ indicates the presence of two gap
values, as observed in experiments\cite{Matano2008}.

%%%%%%%%%%%%%%%%%%%%%%%%%%%%%%%%%%%%%%%%%%%%%%%%%%%%%%%%%%%%%
%\section{3d spin fluctuation}
%%%%%%%%%%%%%%%%%%%%%%%%%%%%%%%%%%%%%%%%%%%%%%%%%%%%%%%%%%%%%
 %%%%%%%%%%%%%%%%%%%%%%%%%%%%%%%%%%%%%%%%%%%%%%%%%%%%%%%%%%%%%
\begin{figure}[t]
\includegraphics{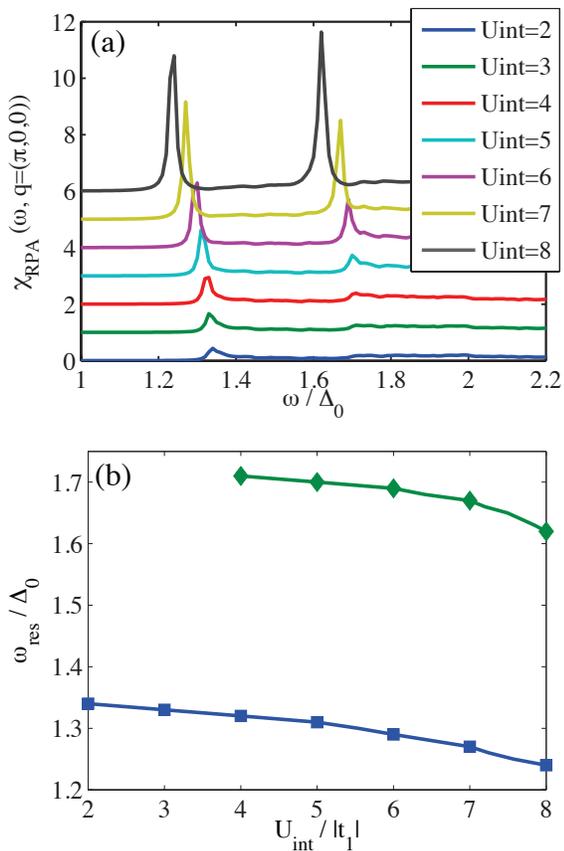}
\caption{ \label{fig5} (a) The RPA spin susceptibilities in the
electron-doped $s_{x^2y^2}$ superconducting  state($\mu=1.65|t_1|$)
with a various $U_{\text{int}}(\mathbf{q})$, and (b) the dispersion
of $\omega_{\text{res}}$ as a function of
$U_{\text{int}}(\mathbf{q})$. }
\end{figure}
%%%%%%%%%%%%%%%%%%%%%%%%%%%%%%%%%%%%%%%%%%%%%%%%%%%%%%%%%%%%%

\paragraph*{The effect of magnetic exchange coupling along c-axis:}
Now we turn to the effects of the spin fluctuations between the FeAs
layers.  For the 122 series of iron-based superconductors, the
coupling between FeAs layers along c-axis is very important. In the
parent compounds, the magnetic exchange coupling $SJ_z$ along c-axis
determined by neutron scattering experiments is around 6 meV - around one fifth of the in-plane magnetic exchange
coupling\cite{Zhao2008c}; this suggests a strong three-dimensional
electronic structure. The measurement of spin excitation resonance
peaks in BaFe$_{1.9}$Ni$_{0.1}$As$_2$ shows they are dispersive along
$c$-axis direction, and the resonance peak at 3D AF ordering
wavevector $(\pi,0,\pi)$ is $\hbar \omega =7$meV below $T_c$
different from one at $\mathbf{Q}_1=(\pi,0,0)$, $\hbar \omega
=9.1$meV\cite{Chi2008}.

  The Hamiltonian for the antiferromagnetic exchange  coupling in c-axis  can be generally written
  as
\begin{eqnarray}\nonumber
H_{\text{int}} &=&  J_{z}\sum_{\mathbf{r},\alpha} \mathbf{S}^\alpha(\mathbf{r})\cdot \mathbf{S}^\alpha(\mathbf{r}\pm \hat{z}) \\
&+& J'_{z}\sum_{\mathbf{r},\alpha \neq \beta} \mathbf{S}^\alpha(\mathbf{r})\cdot \mathbf{S}^\beta(\mathbf{r}\pm \hat{z}),
\end{eqnarray}
%\begin{equation}
%H_{0} = \sum_{\mathbf{k}} \Psi_\mathbf{k}^\dagger
%\left( \begin{array}{cccc}
%\xi_x(\mathbf{k})& \Delta_1(\mathbf{k}) &\epsilon_{xy}(\mathbf{k})& 0 \\
%\Delta_1^\ast(\mathbf{k})& -\xi_x(\mathbf{k}) &0 &-\epsilon_{xy}(\mathbf{k}) \\
%\epsilon_{xy}(\mathbf{k}) &0& \xi_y(\mathbf{k})& \Delta_2(\mathbf{k})  \\
%0& -\epsilon_{xy}(\mathbf{k})& \Delta_2^\ast(\mathbf{k}) &-\xi_x(\mathbf{k})
%\end{array}\right)\Psi_\mathbf{k},
%\end{equation}
where
$J_{z}$ is a spin coupling constant for the intra-orbital and $J'_{z}$ for inter-orbital between the adjacent FeAs layers.
%The dispersion $\xi(\mathbf{k})$'s are the same in the 2 dimensional space, that is,
%\begin{equation}
%\begin{array}{l}
%\epsilon_{x}(\mathbf{k}) = -2 t_1 \cos k_x-2 t_2 \cos k_y -4t_3 \cos k_x\cos k_y , \\
%\epsilon_{y}(\mathbf{k}) = -2 t_1 \cos k_y-2 t_2 \cos k_x -4t_3 \cos k_x\cos k_y, \\
%\epsilon_{xy}(\mathbf{k}) =  -4t_4 \sin k_x\sin k_y.
%\end{array}
%\end{equation}

In general, there are two possible effects generated by this new spin
coupling. First, it could produce the variation of the
superconducting gap as a function of momentum wave-vector along
c-axis.  This effect has been used to explain the dispersion of
magnetic resonance\cite{Chi2008}. Here we point out that the
explanation is only consistent with the extended $s_{x^2y^2}$ wave.
Let's revisit the analysis given in ref.\cite{Chi2008}. Let
$\Delta^e_0$ and $\Delta^h_0$ denote the superconducting gaps on the
hole and electron pockets respectively (in a pure two dimensional
model).  By considering the antiferromagnetic coupling between
layers,  the gap functions, at the mean-field level, are
naturally modified to $\Delta_e(k_z)=\Delta^0_e+\delta \cos(k_z)$ and
$\Delta_h(k_z)=\Delta^0_h+\delta \cos(k_z)$. For a $S^{\pm}$ pairing
symmetry, $\Delta^e_0 \sim -\Delta^h_0\sim -\Delta_0$. Therefore the
dispersion of the resonance along c-axis is roughly determined by
\begin{eqnarray}
\label{resonance} \hbar\omega(q_z) & \sim & \text{min}(\langle|\Delta_e(k_z)|+
|\Delta_h(k_z+q_z)|\rangle, k_z)\nonumber\\ &\sim& 2\Delta_0-2\delta
|\sin(\frac{q_z}{2})|
\end{eqnarray}
In the above analysis, if we assume the sign of gap does not change
from electron to hole pockets, it is easy to see that the dispersion
of the resonance in Eq.\ref{resonance}  changes to $\hbar\omega(q_z)
\sim 2\Delta_0+2\delta |\sin(\frac{q_z}{2})|$, which will result in
an opposite dispersion of   the resonance energy than that reported
in the experiments:  larger at the wavevector $(\pi,0,\pi)$ than  at
the wavevector $(\pi,0,0)$,
  obviously contradicting the experimental results.

There is also a second effect due to the magnetic exchange coupling
along c-axis, namely, a simple RPA correction to spin susceptibility
due to the presence of the exchange coupling. It is easy to show
that  in the presence of above exchange coupling, the RPA spin
susceptibility is modified to \begin{equation} \chi^{\text{RPA}}
(\mathbf{q},\omega )= \sum_{\alpha\beta} [ \chi_0(\mathbf{q},\omega
) ( 1 - V(\mathbf{q}) \chi_0(\mathbf{q},\omega
))^{-1}]_{\alpha\beta},
\end{equation}
with the vertex
\begin{equation}
V(\mathbf{q})=\left(\begin{array}{cc}
U_{\text{int}}(\mathbf{q}) & -J'_{z}\cos q_z \\
-J'_{z}\cos q_z & U_{\text{int}}(\mathbf{q})
\end{array}\right),
\end{equation}
where $U_{\text{int}}(\mathbf{q}) = U - J_{z}\cos q_z$ with $U$ a
onsite Coulomb repulsion.  We calculate
$\chi_{\text{RPA}}(\omega,\mathbf{q})$ in the electron-doped
superconducting state($\mu=1.65|t_1|$)  with $s_{x^2y^2}$ pairing
symmetry, $\Delta(\mathbf{k}) =\Delta_0 \cos k_x \cos k_y$.
  $J'_{z}$, the inter-orbital coupling has very little effect on the
positions of $\omega_{\text{res}}$. Therefore, we only need to
consider the   intra-orbital spin coupling between the layers,
$J_{z}$. The effect of $J_z$ simply creates an effective
$U_{\text{int}}(q)$ which is larger at $q=(\pi,0,\pi)$ than at
$q=(\pi,0,\pi)$. Fig.~\ref{fig5}(b) displays the dispersion of
$\omega_{\text{res}}$ as a function of $U_{\text{int}}(\mathbf{q})$.
It is clear that the resonance energy decreases as $U_{\text{int}}$
increases. Therefore, in general, the RPA correction would result in
a dispersion of resonance peak that has a correct trend as the
experimental results. However, it is also clear from Fig.\ref{fig5}
that $J_z$ has to be comparable with $U$ in order to create a
visible dispersion. Since $J_z$ is really a small energy scale
compared to both band width and interaction,  the dispersion due to
RPA correction from $J_z$ is negligible.

From above analysis, we can conclude that the dispersion of the
resonance peak along c-axis mainly stems from the modification the
SC gap along c-axis. The fact that the dispersion of the resonance
peak towards lower energy from $Q_z=0$ to $Q_z=\pi$ indicates that
the superconducting order parameters at electron and hole pockets
must have opposite values.

%%%%%%%%%%%%%%%%%%%%%%%%%%%%%%%%%%%%%%%%%%%%%%%%%%%%%%%%%%%%%
\paragraph*{Conclusion}
%%%%%%%%%%%%%%%%%%%%%%%%%%%%%%%%%%%%%%%%%%%%%%%%%%%%%%%%%%%%%
We have  investigated the magnetic properties in the SC states of
the iron-based superconductors based on a two-orbital model,
identifying generic, model-independent, experimentally observable
predictions.    We have found the
 $s_{x^2y^2}$ pairing symmetry exhibits a strong magnetic resonance, which is absent from other
 pairing symmetries(the similar result was also reached in \cite{Maier2009} when this paper was prepared).
    We have predicted that  the
coexistence of SDW order with the SC order weakens the resonance
peak. We have shown a good agreement with NMR experimental results,
both the $1/T_1$ and the Knight shift,  can be reached after
considering the RPA correction from a bare BCS state with the
extended $s_{x^2y^2}$ pairing symmetry. We have also explicitly
studied the dispersion of magnetic resonance along
c-axis\cite{Chi2008} and have concluded that the dispersion is
mainly caused by the modification of SC gap along c-axis and  it
indicates that the superconducting order parameters at electron and
hole pockets must have opposite values. This explanation is also
consistent with the three-dimensional nature of the 122 series of
iron-pnicitides as shown in different
experiments\cite{Zhao2008c,Yuan2009}. Finally, we also want to point
out that this explanation does not contradict the possible existence
of  gap nodes along c axis which may explain many nodal-like
experimental results in the 122 materials\cite{Check2008,Gordon2008}
since the resonance peak is determined by the summation of the
absolute values of SC gaps at electron and hole pockets.

\paragraph*{Acknowledge} JPH thanks  Pengcheng Dai,  W. Tsai, and DX Yao for useful
discussions. BAB thanks P. W. Anderson and N. P. Ong for useful
discussions. JPH, KJS and CF were supported by the NSF under grant
No. PHY-0603759.
\bibliographystyle{prsty}
\bibliography{iron3}

\end{document}